\documentclass[12pt]{cernrep}
\usepackage{graphicx}
\usepackage{here}

\def \beq{\begin{equation}}

\def \eeq{\end{equation}}

\def \rpp{R_{\pi \pi}}

\def \beqa{\begin{eqnarray}}
\def \eeqa{\end{eqnarray}}
\begin{document}
 \title{\large Determination of weak phases $\phi_2$ and $\phi_3$ \\
from $B\to \pi\pi,K\pi$ in the pQCD method}
\author{ {\bf Yong-Yeon Keum} \\
EKEN Lab. Department of Physics \\
Nagoya University, Nagoya 464-8602 Japan}
\maketitle
\begin{abstract}
We look at methods to determine the weak phases
$\phi_2$ and $\phi_3$ from $B \to \pi\pi$ and $K\pi$ decays within
the perturbative QCD approach. We obtain quite interesting bounds
on $\phi_2$ and $\phi_3$ from recent experimental measurement in B-factory:
$55^o \leq \phi_2 \leq 100^o$ and $51^o \leq \phi_3 \leq 129^o$.
Specially we predict the possibility of large direct CP violation
effect in $B^0 \to \pi^{+}\pi^{-}\,(23\pm7 \%)$ decay.

\end{abstract}
\vspace{-2.5in}
\rightline{DPNU-02-28} 
\rightline{hep-ph/0209002}
\rightline{Presented at CERN CKM Workshop}
\rightline{13--16 February 2002}
\vspace{2in}

\section{INTRODUCTION}
One of the most exciting aspect of present high energy physics is 
the exploration of CP violation in B-meson decays,
allowing us to overconstrain both sides and three weak phases
$\phi_1(=\beta)$, $\phi_2(=\alpha)$ and $\phi_3(=\gamma)$ of the
unitarity triangle of the Cabibbo-Kobayashi-Maskawa (CKM) matrix
\cite{ckm} and to check the possibility of New Physics.
The ``gold-plated'' mode $B_d \to J/\psi K_s$\cite{sanda}, 
which allow us to determine $\phi_1$ without any hadron uncertainty,
was recently measured by BaBar and Belle collaborations\cite{bfactory}:
$\phi_1=(25.5\pm4.0)^o$. In addition,
there are many other interesting channels with which we may achieve this
goal by determining $\phi_2$ and $\phi_3$\cite{gamma}.

In this letter, we focus on the $B \to \pi^{+}\pi^{-}$ and
$K\pi$ processes, providing promising strategies for determining
the weak phases of $\phi_2$ and $\phi_3$, 
by using the perturbative QCD method.

The perturbative QCD method (pQCD) has 
predictive power demonstrated successfully 
in exclusive 2 body B-meson decays,
specially in charmless B-meson decay processes\cite{pQCD}. 
By introducing parton transverse momenta $k_{\bot}$, 
we can generate naturally the Sudakov suppression effect 
due to resummation of large double 
logarithms $Exp[-{\alpha_s C_F \over 4 \pi} \ln^2({Q^2\over k_{\bot}^2})]$,
which suppress the long-distance contributions in the small $k_{\bot}$ region
and give a sizable average $<k_{\bot}^2> \sim \bar{\Lambda} M_B$. 
This can resolve the end point singularity problem and 
allow the applicability of pQCD to exclusive decays. We found that
almost all of the contribution to the exclusive matrix elements
come from the integration region where $\alpha_s/\pi < 0.3$ and 
the pertubative treatment can be justified.

In the pQCD approach, we can predict the contribution of non-factorizable
term and annihilation diagram on the same basis
 as the factorizable one.
A folklore for annihilation contributions is that they are negligible
compared to W-emission diagrams due to helicity suppression. 
However the operators $O_{5,6}$ with helicity structure $(S-P)(S+P)$
are not suppressed and give dominant imaginary values, 
which is the main source of strong phase in the pQCD approach.
So we have a large direct CP violation in 
$B \to \pi^{\pm}\pi^{\mp}, K^{\pm}\pi^{\mp}$,
since large strong phase comes from 
the factorized annihilation diagram, which can distinguish pQCD from
other models\cite{bbns,charm}.  

\begin{figure}[ht]
\vskip-1.0cm
\begin{center}
\includegraphics[angle=-90,width=12.0cm]{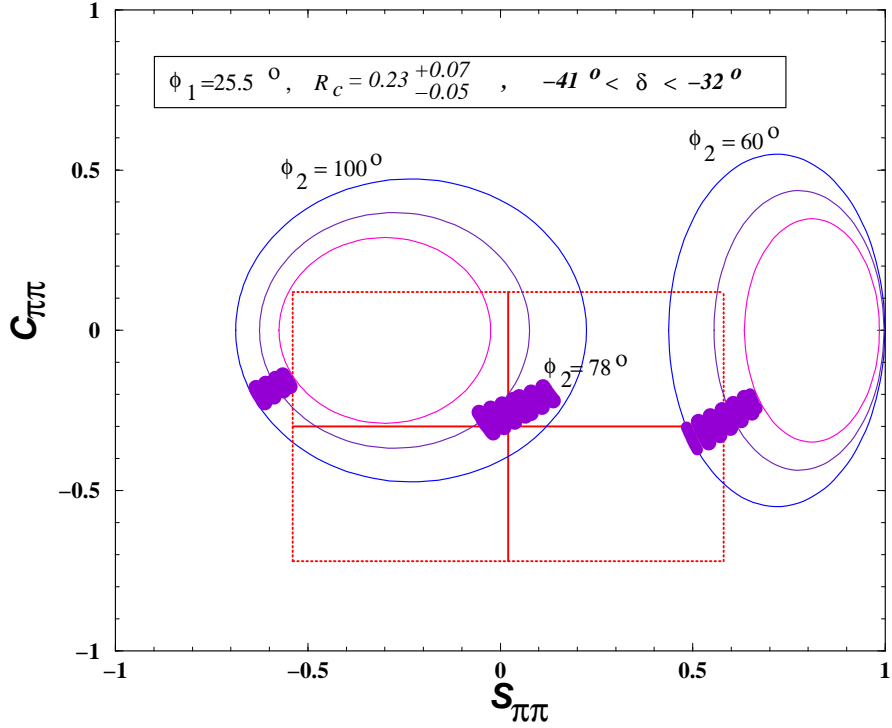} 
\caption{Plot of $C_{\pi\pi}$ versus $S_{\pi\pi}$  for various values
of $\phi_2$ with $\phi_1=25.5^o$, $0.18 < R_c < 0.30$ and $-41^o <
\delta < -32^o$ in the pQCD method. Here we consider the allowed experimental
ranges of BaBar measurment whinin $90\%$ C.L. 
Dark areas is allowed
regions in the pQCD method for different $\phi_2$ values.}
\end{center}
\label{fig:cpipi}
\end{figure}

\section{Extraction of $\phi_2(=\alpha)$ from $B \to \pi^{+}\pi^{-}$}
Even though isospin analysis of $B \to \pi\pi$ can provide a clean way
to determine $\phi_2$, it might be difficult in practice because of
the small branching ratio of $B^0 \to \pi^0\pi^0$.
In reality to determine $\phi_2$, we can use the time-dependent rate
of $B^0(t) \to \pi^{+}\pi^{-}$ including sizable penguin
contributions. In our analysis we use the c-convention.
The amplitude can be written as:
\beqa
A(B^0\to \pi^{+}\pi^{-}) &=& V_{ub}^{*}V_{ud} A_u + V_{cb}^{*}V_{cd} A_c
+ V_{tb}^{*}V_{td} A_t,\nonumber \\
&=& V_{ub}^{*}V_{ud}\,\, (A_u-A_t) + V_{cb}^{*}V_{cd} (A_c-A_t),
\nonumber \\
&=& -(|T_c|\,\,e^{i\delta_T} \, e^{i\phi_3} + |P_c|\, e^{i\delta_P})
\eeqa
Pengun term carries a different weak phase than the dominant tree amplitude,
which leads to generalized form of the time-dependent asymmetry:
\beq
 A(t) \equiv {\Gamma(\bar{B}^0(t) \to \pi^{+}\pi^{-}) - 
\Gamma(B^0(t) \to \pi^{+}\pi^{-}) \over \Gamma(\bar{B}^0(t) \to \pi^{+}\pi^{-}) + 
\Gamma(B^0(t) \to \pi^{+}\pi^{-})} = 
S_{\pi\pi} \,\,sin(\Delta m t) - C_{\pi\pi}\,\, cos(\Delta m t)
\eeq
where 
\beq
C_{\pi\pi}={1-|\lambda_{\pi\pi}|^2 \over 1+|\lambda_{\pi\pi}|^2},
\hspace{20mm}
S_{\pi\pi}={2 \,Im(\lambda_{\pi\pi}) \over 1+|\lambda_{\pi\pi}|^2}
\eeq
satisfies the relation of $C_{\pi\pi}^2 + S_{\pi\pi}^2 \leq 1$.
Here 
\beq
\lambda_{\pi\pi} = |\lambda_{\pi\pi}|\, e^{2i(\phi_2 + \Delta\phi_2)}
=e^{2i\phi_2} \left[{1+R_c e^{i\delta} \,e^{i\phi_3} \over 
1+R_c e^{i\delta} \,e^{-i\phi_3} } \right]
\eeq
with $R_c=|P_c/T_c|$ and the strong phase difference
between penguin and tree amplitudes $\delta=\delta_P-\delta_T$.
The time-dependent asymmetry measurement provides two equations for
$C_{\pi\pi}$ and $S_{\pi\pi}$ 
for three (unknown) variables $R_c,\delta$ and $\phi_2$.

When we define $\rpp=\overline{Br}(B^0 \to \pi^{+}\pi^{-})/
\overline{Br}(B^0\to \pi^{+}\pi^{-})|_{tree}$, 
where $\overline{Br}$ stands for 
a branching ratio averaged over $B^0$ and $\bar{B}^0$, the explicit
expression for $S_{\pi\pi}$ and $C_{\pi\pi}$ are given by:
\beqa
R_{\pi\pi} &=& 1-2\,R_c\, cos\delta \, cos(\phi_1 +\phi_2) + R_c^2,  \\
R_{\pi\pi}S_{\pi\pi} &=& sin2\phi_2 + 2\, R_c \,sin(\phi_1-\phi_2) \,
cos\delta - R_c^2 sin2\phi_1, \\
R_{\pi\pi}C_{\pi\pi} &=& 2\, R_c\, sin(\phi_1+\phi_2)\, sin\delta.
\eeqa
If we know $R_c$ and $\delta$, $\phi_2$ can be determined from the
experimental data on $C_{\pi\pi}$ versus $S_{\pi\pi}$. 

Since the pQCD method provides $R_c=0.23^{+0.07}_{-0.05}$ and $-41^o
<\delta<-32^o$, the allowed range of $\phi_2$ at present stage is
determined as $55^o <\phi_2< 100^o$ as shown in Fig. \ref{fig:cpipi}. 
Since we have a relatively large
strong phase in pQCD, 
in contrast to the QCD-factorization ($\delta\sim 0^o$), we predict
large direct CP violation effect of 
$A_{cp}(B^0 \to \pi^{+}\pi^{-}) = (23\pm7) \%$ which will be tested
by more precise experimental measurement within 3 years. 
In numerical analysis, since the data by Belle
collaboration\cite{belle} 
is located outside allowed physical regions, we only considered the
recent BaBar measurement\cite{babar} with $90\%$ C.L. interval
taking into account the systematic errors:
\begin{itemize}
\item[$\bullet$]
$S_{\pi\pi}= \,\,\,\,\, 0.02\pm0.34\pm0.05$ 
\hspace{10mm} [-0.54,\hspace{5mm} +0.58]
\item[$\bullet$]
$C_{\pi\pi}=-0.30\pm0.25\pm0.04$ 
\hspace{10mm} [-0.72,\hspace{5mm} +0.12].
\end{itemize}
The central point of BaBar data corresponds to $\phi_2 = 78^o$ 
in the pQCD method. 

\begin{figure}[ht]
\begin{center}
\includegraphics[angle=0,width=12.0cm]{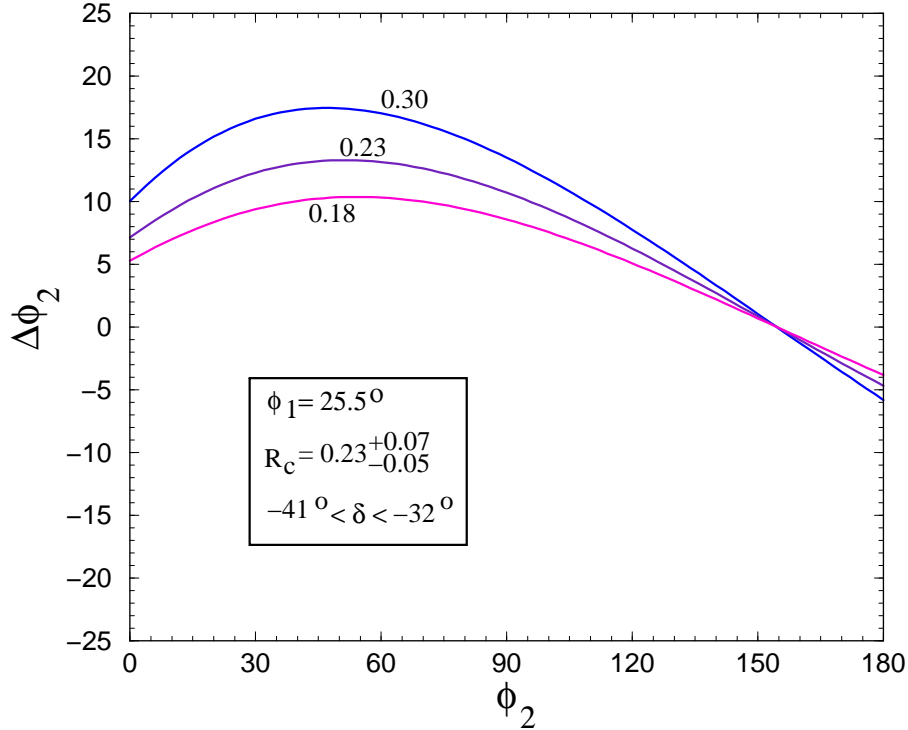} 
\caption{Plot of $\Delta \phi_2$ versus $\phi_2$ with
$\phi_1=25.5^o$, $0.18 < R_c < 0.30$ and $-41^o <
\delta < -32^o$ in the pQCD method.}
\end{center}
\label{fig:delphi2}
\end{figure}

Denoting by $\Delta \phi_2$ the deviation of $\phi_2$ due to the penguin
contribution, derived from Eq.(4), 
it can be determined for known values of $R_c$ and $\delta$ by using
the relation $\phi_3 = 180 -\phi_1 -\phi_2$. 
In Fig. \ref{fig:delphi2} we show that
pQCD prediction on the relation $\Delta\phi_2$ versus $\phi_2$.
For allowed regions of $\phi_2=(55 \sim 100)^o$, 
we have $\Delta \phi_2 =(8
\sim 16)^o$. The main uncertainties come from the uncertainty of
$|V_{ub}|$. The non-zero value of $\Delta \phi_2$ for $55^o < \phi_2 < 100^o$
demonstrates sizable
penguin contributions in $B^0 \to \pi^{+}\pi^{-}$ decay. 

\section{Extraction of $\phi_3(=\gamma)$ 
from $B^0 \to K^{+}\pi^{-}$ and $B^{+}\to K^0\pi^{+}$ Processes}
By using tree-penguin interference in $B^0\to K^{+}\pi^{-}(\sim
T^{'}+P^{'})$ versus $B^{+}\to K^0\pi^{+}(\sim P^{'})$, CP-averaged
$B\to K\pi$ branching fraction may lead to non-trivial constaints
on the $\phi_3$ angle\cite{fle-man}. In order to determine $\phi_3$,
we need one more useful information 
on CP-violating rate differences\cite{gr-rs02}.
Let's introduce the following observables :
\beqa
R_K &=&{\overline{Br}(B^0\to K^{+}\pi^{-}) \,\, \tau_{+} \over
\overline{Br}(B^+\to K^{0}\pi^{+}) \,\, \tau_{0} }
= 1 -2\,\, r_K \, cos\delta \, \, cos\phi_3 + r_K^2 \nonumber \\
&& \hspace{40mm} \geq sin^2\phi_3     \\
\cr
A_0 &=&{\Gamma(\bar{B}^0 \to K^{-}\pi^{+} - \Gamma(B^0 \to
K^{+}\pi^{-}) \over \Gamma(B^{-}\to \bar{K}^0\pi^{-}) +
 \Gamma(B^{+}\to \bar{K}^0\pi^{+}) } \nonumber \\
&=& A_{cp}(B^0 \to K^{+}\pi^{-}) \,\, R_K = -2 r_K \, sin\phi_3 \,sin\delta.
\eeqa
where $r_K = |T^{'}/P^{'}|$ is the ratio of tree to penguin amplitudes
and $\delta = \delta_{T'} -\delta_{P'}$ is the strong phase difference
between tree and penguin amplitides.
After eliminate $sin\delta$ in Eq.(8)-(9), we have
\beq
R_K = 1 + r_K^2 \pm \sqrt(4 r_K^2 cos^2\phi_3 -A_0^2 cot^2\phi_3).
\eeq
Here we obtain $r_K = 0.201\pm 0.037$ 
from the pQCD analysis\cite{pQCD} 
and $A_0=-0.110\pm 0.065$ by combining recent BaBar
measurement on CP asymmetry of $B^0\to K^+\pi^-$: 
$A_{cp}(B^0\to K^+\pi^-)=-10.2\pm5.0\pm1.6 \%$ \cite{babar}
with present world averaged value of  $R_K=1.10\pm 0.15$\cite{rk}.

\begin{figure}[ht]
\begin{center}
\includegraphics[angle=-90,width=12.0cm]{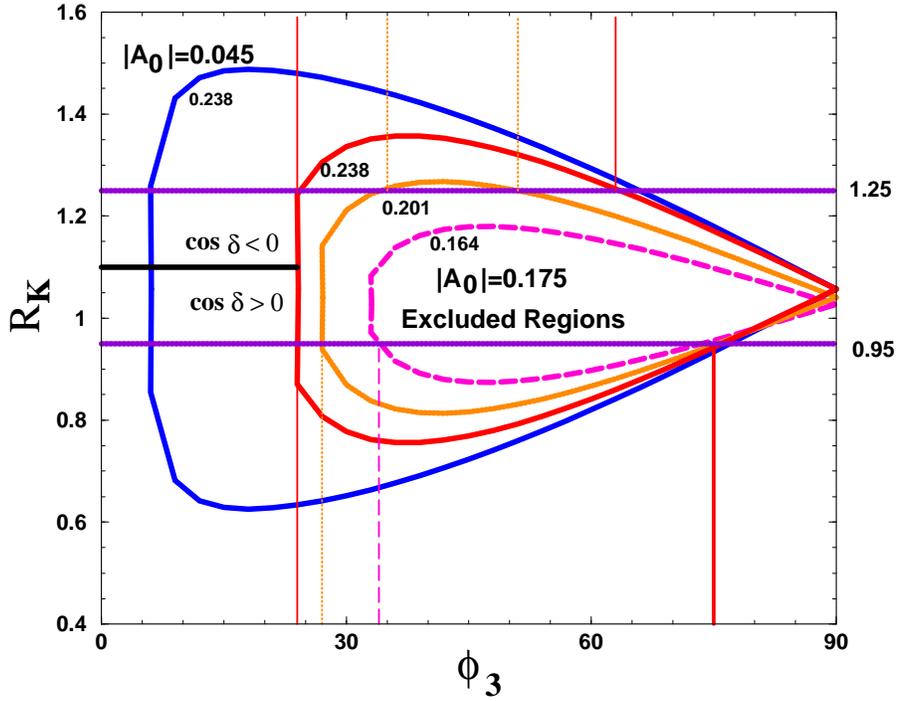} 
\caption{Plot of $R_K$ versus $\phi_3$ with $r_K=0.164,0.201$ and $0.238$.}
\end{center}
\label{fig:Rkpi}
\end{figure}
As shown in Fig.\ref{fig:Rkpi}, we can constrain the allowded $\phi_3$ 
with $1\,\sigma$ range of World Averaged $R_K$ as follows:
\begin{itemize}
\item[$\bullet$]For $cos\delta > 0$, $r_K=0.164$: we can exclude
$0^o \leq \phi_3 \leq 6^0$ and $ 24^o \leq \phi_3 \leq 75^0$. 
\item[$\bullet$]For $cos\delta > 0$, $r_K=0.201$: we can exclude
$0^o \leq \phi_3 \leq 6^0$ and $ 27^o \leq \phi_3 \leq 75^0$. 
\item[$\bullet$]For $cos\delta > 0$, $r_K=0.238$: we can exclude
$0^o \leq \phi_3 \leq 6^0$ and $ 34^o \leq \phi_3 \leq 75^0$.
\item[$\bullet$]For $cos\delta < 0$, $r_K=0.164$: we can exclude
$0^o \leq \phi_3 \leq 6^0$. 
\item[$\bullet$]For $cos\delta < 0$, $r_K=0.201$: we can exclude
$0^o \leq \phi_3 \leq 6^0$ and $ 35^o \leq \phi_3 \leq 51^0$. 
\item[$\bullet$]For $cos\delta < 0$, $r_K=0.238$: we can exclude
$0^o \leq \phi_3 \leq 6^0$ and $ 24^o \leq \phi_3 \leq 62^0$.
\end{itemize}

From the table 2 of ref.\cite{keum},
we obtain $\delta_{P'} = 157^o$, $\delta_{T'} = 1.4^o$, and
the negative value of $cos\delta$: $cos\delta= -0.91$.
The maximum value of the constraint bound for the $\phi_3$
strongly depends on the value of $|V_{ub}|$.
When we take the central value of $r_K=0.201$,
$\phi_3$ is allowed within the ranges of $51^o \leq \phi_3 \leq
129^o$, which is consistent with the results by the model-independent
CKM-fit in the $(\rho,\eta)$ plane.

\section{CONCLUSION}
We discussed two methods to determine the weak phases 
$\phi_2$ and $\phi_3$ within the pQCD
approach through 1) Time-dependent asymmetries in $B^0\to
\pi^{+}\pi^{-}$, 2) $B\to K\pi$ processes via penguin-tree
interference. We can already obtain interesting bounds on $\phi_2$
and $\phi_3$ from present experimental measurements.
Our predictions within pQCD method is in good agreement with present
experimental measurements in charmless B-decays.
Specially our pQCD method predicted a large direct CP asymmetry
in $B^0 \to \pi^{+}\pi^{-}$ decay, which will be a crucial touch stone
in order to distinguish our approach from others in future precise measurements.
More detail works on other methods in $B\to K\pi$ and $D^{(*)}\pi$ processes
will appeare elsewhere. \cite{keum-02}. 
\vskip1cm
\noindent

\section*{ACKNOWLEDGEMENTS}
It is a great pleasure to thank E.~Paschos,
A.I. Sanda, H.-n. Li, and other members of PQCD working group for
fruitful collaborations and joyful discussions.
We would like to thank S.J.~Brodsky, H.Y.~Cheng and M.~Kobayashi for their
hospitality and encouragement. This work was supported in part by
Science Council of R.O.C. under Grant No. NSC-90-2811-M-002 and
in part by Grant-in Aid for Scientific Reserach from Ministry of
Education, Science and Culture of Japan.

\end{document}